\documentclass[12pt,preprint]{aastex}

\begin{document}

\title{The inner annular Gap for pulsar radiation:
$\gamma$-ray and radio emission}

\author{G.J. Qiao$^{1,2}$, K.J. Lee$^2$, H.G. Wang$^3$, R.X. Xu$^2$ \& J.L.
Han$^{1,2}$}

\affil{ $^1$National Astronomical Observatories, Chinise Academy
of Sciences, Beijing 100012, China\\
$^2$ Department of Astronomy, Peking University, Beijing 100871, China\\
$^3$ Center for Astrophysics, Guangzhou University, Guangzhou
510400}

\begin{abstract}
The inner annular gap (IAG), a new type of inner gap whose
magnetic field lines intersect the null charge surface (NCS), is
proposed to explain $\gamma$-ray and radio emission from pulsars.
The IAG can be an important source for high-energy particles. The
particles can radiate between the NCS and the IAG. Some
observational characteristics in both $\gamma$-ray and radio
bands, such as the $\gamma$-ray emission beams of Crab-like,
Vela-like and Geminga-like, can be reproduced by numerical method.
It is predicted that the view angle $\zeta$ should be larger than
the inclination angle ($\zeta>\alpha$), otherwise the
$\gamma$-ray radiation will have little possibility to be
observed. Whether the inner annular gap (or cap) is sparking (or
free flow) depends on the surface binding energy of the pulsar. In
stead of neutron star models, the scenario of the IAG is favorable
for bare strange star models, because bare strange stars can
easily satisfy the requisite condition to form an IAG for both
pulsars ($\vec{\Omega}\cdot \vec{B}<0$) and anti-pulsars
($\vec\Omega \cdot \vec{B}>0$).

\end{abstract}


\section{Introduction}

\begin{figure}
\plotone{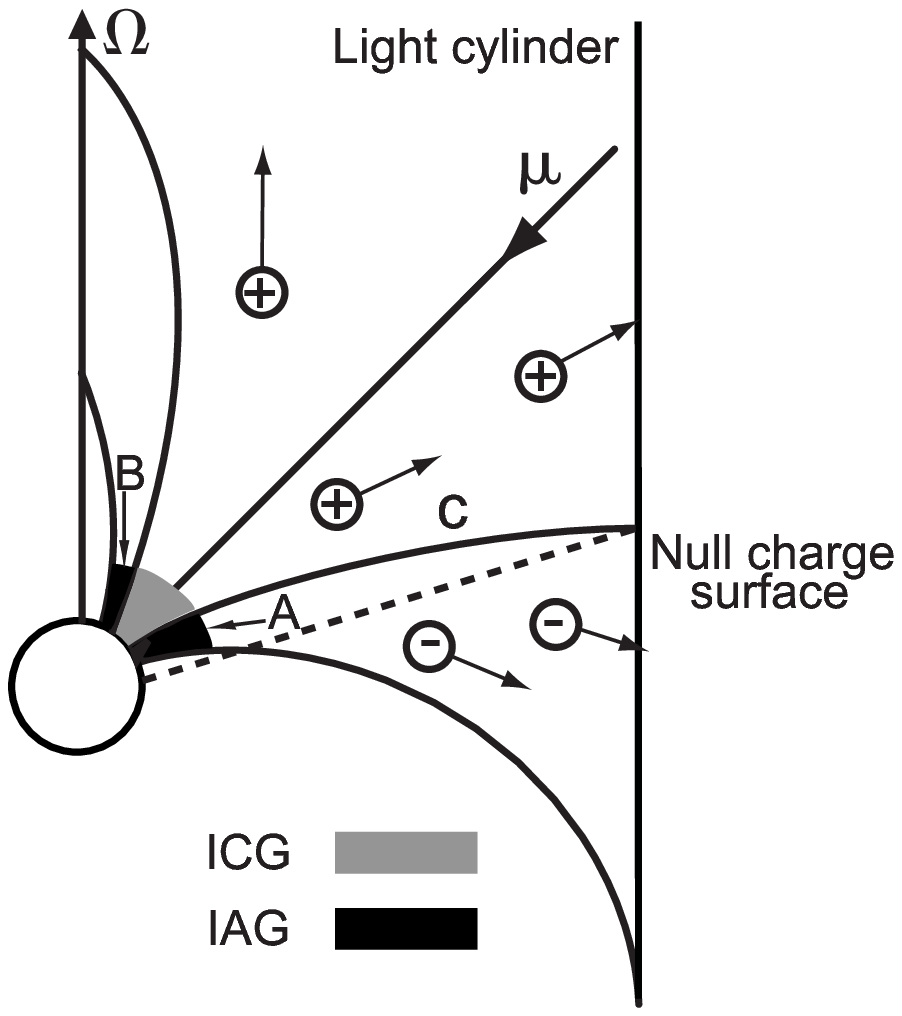} \caption{The inner annular gap (IAG), the inner
core gap (ICG) and the null charge surface (NCS) of an oblique
rotator. If the center star is a strange star, both IAG and ICG
can form. For neutron stars, at most only one inner gap can be
formed. Region 'A' and 'B' correspond to the region 'A' and 'B' on
Fig.~\ref{cap}. } \label{GJcg}
\end{figure}

Both polar cap and outer gap models of $\gamma$-ray pulsars have
been suggested to explain the high energy emission from pulsars
 \citep{A81,H81,HM98,MH03,Z89,L90,L94,C76,R81,CHR86,C00,H00,R96,R02}.
Here, we propose a new scenario of the inner annular gap
(hereafter IAG) outside the conventional inner gap. The IAG is
defined by magnetic field lines which intersect the null charge
surface(NCS) (see Fig.\ref{GJcg} ). As we know, at any point on
the NCS the direction of local magnetic field is perpendicular to
the rotation axis. So if the radiation region locates near the NCS
(either inside or outside), the pulsar can easily produce a very
wide pulse profile \citep{Q03a}. This wide profile will match the
observed $\gamma$-ray pulse profiles rather well. Some authors
\citep{Q03a,Q03b,DR03} suggested that if the emission region
locates inside the NCS, it would be favorable to understand the
$\gamma$-ray emission. Here we suggest that the particles flow out
from the IAG could be re-accelerated and radiate $\gamma$-rays
inside the NCS, and the model to account for the $\gamma$-ray and
radio emission at the same time. In \S 2 the basic ideas for the
IAG are presented. In \S 3 we show how $\gamma$-ray and radio
emissions can be produced. Conclusions and brief discussions are
given in \S 4.

\section{Details of the inner annular gap (IAG)}

Fig.~\ref{GJcg} illustrates the magnetosphere of an oblique
rotator with a dipolar magnetic field configuration. The magnetic
field line 'c' is the \emph{critical field line}, which intersects
the NCS at the light cylinder. The radius of the polar cap region
defined by the critical field lines in an aligned rotator is
$r_{\rm in}=0.74\Omega^{0.5} R^{1.5} c^{0.5}$ \citep{RS75}. Here
$R$ and $\Omega$ are the radius and the angular velocity of the
star, respectively, and $c$ is the light speed. The cap radius to
the foot of the last open field line is $r_{\rm p}= \Omega^{0.5}
R^{1.5} c^{0.5}$. There is an annulus between $r_{\rm p}$ and
$r_{\rm in}$.

Two kinds of inner vacuum gaps above the polar cap may be formed
in some circumstances: the conventional \emph{inner core gap}
(ICG) above the central part of the polar cap, and the \emph{inner
annular gap} (IAG) above the annular part of the polar cap. If the
width of IAG is large enough, the potential drop in the IAG would
be high enough so that sparking will be able to take place there.
The sparking leads to pair production and generates the secondary
pairs, which are accelerated out of the IAG.

\begin{figure}
\plotone{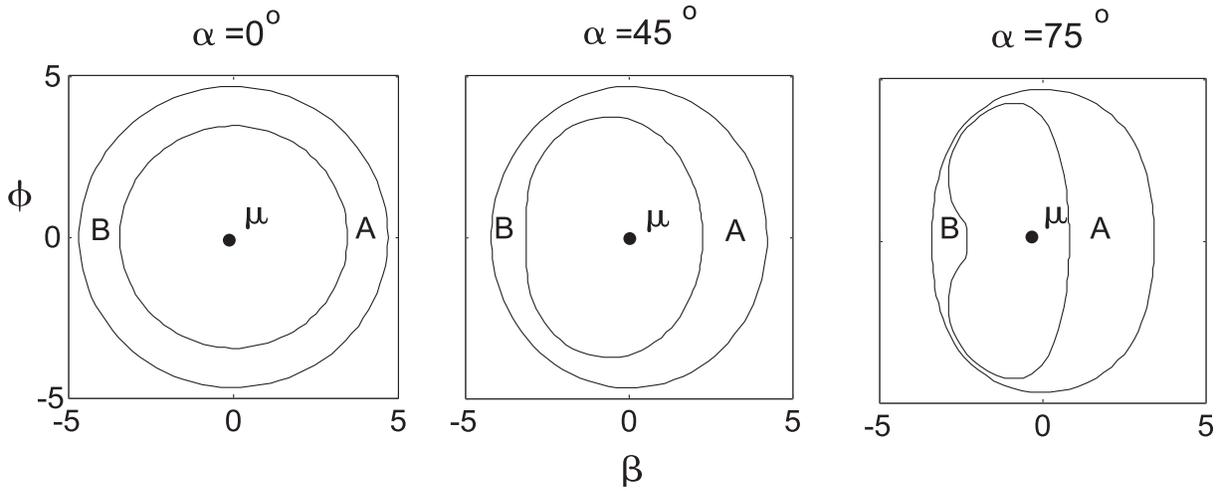} \caption {The shape of the inner annular gap(IAG)
and the inner core gap(ICG) for different inclination angle
$\alpha$. A dipole configuration is used in the calculations. We
take radius of light cylinder $R_{\rm cl}=1500$km and the radius
of pulsar $R=10$km.} \label{cap}
\end{figure}

For neutron stars, the binding energy of positive particles could
be high enough only under some special conditions \citep{GM01,G02}
to lead to the formation of an inner vacuum gap. Only one of the
inner vacuum gaps can form in the case: an IAG for pulsar
($\vec{\Omega}\cdot \vec{B}<0$) or an ICG for anti-pulsar
($\vec{\Omega} \cdot \vec{B}>0$). However the physical condition
changes for bare strange stars. The IAG and ICG can form in this
case no matter the star is a pulsar or an anti-pulsar since the
binding energy is roughly infinite \citep{X99}. The width of IAG
is a function of the inclination angle $\alpha$, as Fig.~\ref{cap}
shows. In the dipolar configuration, the field lines are traced to
obtain the shapes of IAG and ICG. When $\alpha$ increases, the
width of the annular gap between the magnetic axis and the equator
(region A), becomes wider. Higher potential drop can be obtained
to produce sparks. While the width of IAG between the rotational
and magnetic axes (region B) becomes narrow, and has little chance
to form sparks.

Pair cascades can develop in the IAG, and the secondary pairs are
produced in an intrinsically non-stationary mode, just as what
occur in RS type gap. The secondary pairs are generated as small
bundles. In RS gap model these bundles of non-neutral plasma
endure no more acceleration out of the gap. Does it still hold
true for the IAG scenario? It is probably not, we note that
several factors may lead to re-acceleration out of the IAG.

Firstly, due to field line curvature, the net charge density will
depart from the local GJ charge density when particles moves along
the magnetic field lines, which induces parallel electric fields.
This effect is especially strong in the magnetic tube rooted in
the IAG, because the GJ charge density varies more sharply than on
other open field lines and even the local changes from one sign to
the opposite sign when moving across the NCS. Secondly, it was
normally suggested that the self-consistent adjustment of the net
charge of non-neutral plasma is able to screen the parallel
electric fields. However, when charged particles are moving at
relativistic velocities and are bunched into bundles, one has to
calculate the electric field consulting the Li\'{e}nard-Wiechert
potential rather than the Coulomb potential. It is known that for
a relativistic charge particle (with a Lorentz factor $\gamma$)
the electric field on the moving direction is reduced by a factor
of $1/\gamma^2$, then the screen electric field generated by the
secondary pairs should be also reduced by a factor of
$1/\gamma^{2}$, which may be neglected to compare with the
re-acceleration electric field \citep{L03}.

In this paper we simply assumed that the secondaries are
re-accelerated and radiate photons in the region between the NCS
and the IAG, then we aim to find whether it is feasible to account
for $\gamma$-ray and radio properties of pulsar emission. The next
section is an endeavor to reproduce the observational properties
at $\gamma$-ray and radio bands based on the assumption.

\section{$\gamma$-ray and radio emission from pulsars}

\begin{figure}
\plotone{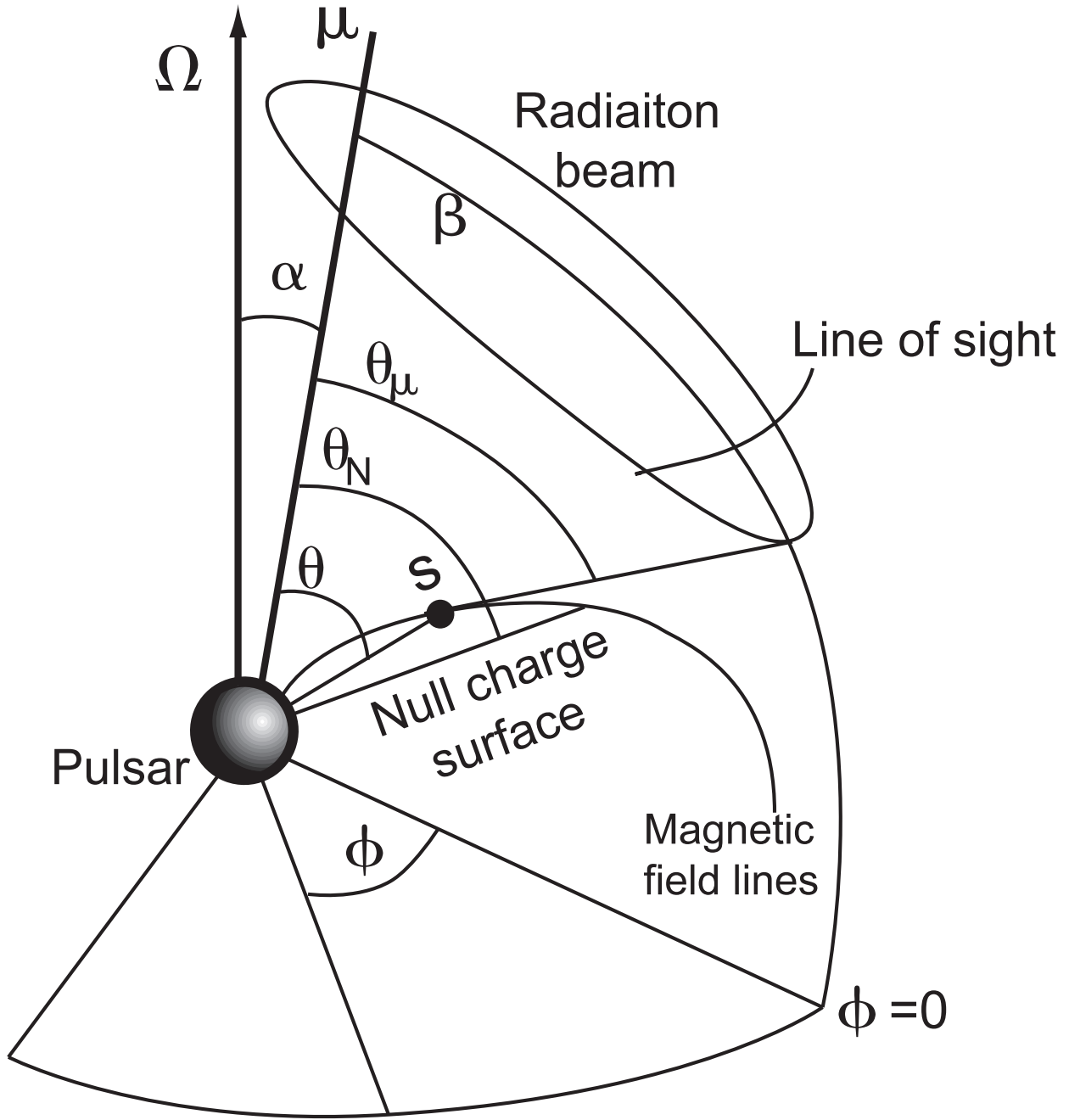} \caption{The radiation geometry of a pulsar.
$\phi$, $\zeta$ and $\alpha$ are the azimuthal angle respect to
magnetic axis, the viewing angle and the inclination angle,
respectively. $\theta$, $\theta_{\rm \mu}$ and $\theta_{\rm N}$
denote the angle between the magnetic axis and the radiation
source $S$, the radiation direction and the null charge surface
respectively. Owing to the aberration effect and the asymmetry of
the NCS respect
 of magnetic axis, the radiation beam is asymmetric.} \label{geo}
\end{figure}

{\em Basic picture for $\gamma$-ray emission}. As it is assumed
that $\gamma$-ray emission can be produced between NCS and the
IAG, also wide pulse profiles can be produced there. To simulate
the observations geometry is important. The radiation geometry is
shown in Fig.~\ref{GJcg} \& ~\ref{geo}, where $\phi$ is the
azimuthal angle around the magnetic axis, $\alpha$ is its
inclination angles, $\theta$ is the angle between the magnetic
axis and the radiation location $r$. The angle between the
magnetic axis and the NCS, $\theta_{\rm N}$, has the form of

\begin{equation}
\theta_{\rm N}(\phi)=\frac{1}{2}\arccos
\left[\frac{\chi\cos\phi\sin\alpha-\cos^{2}\alpha}{3(\cos^{2}\alpha+
\cos^{2}\phi\sin^{2}\alpha)}\right],
\end{equation}

where $\chi=(8\cos^{2}\alpha+ 9\cos^{2}\phi\sin^{2}\alpha)^{0.5}$.
The distance from the center of the star to the intersection of
last open magnetic field lines and the NCS is denoted by $r_{\rm
N}(\phi)$. For a dipole field configuration, $r_{\rm
N}(\phi)=R_{\rm 0}(\phi) \sin^{2}[\theta_{\rm N}(\phi)]$, where
$R_{\rm 0}(\phi)$ is the maximum radius of the last open field
line. The distance from the star center to the emission point is
$r(\phi)=\kappa (\lambda  r_{\rm N}(\phi)+(1-\lambda) r_{\rm
N}(0))$, where $\lambda$ and $\kappa$ are two parameters used to
indicate the radiation location. For simplicity, we take $\lambda$
and $\kappa$ as 0.8 in our calculations. The physical meaning of
$\kappa=0.8 $ is that the radiation positions locate near the NCS
and the distance of the radiation location is roughly in
proportion to the distance of NCS. $\lambda=0.8$ means that the
distance of radiation location is no exactly in proportion to that
of NCS and needs correction, when the NCS locates far form the
star.

\begin{figure}
\plotone{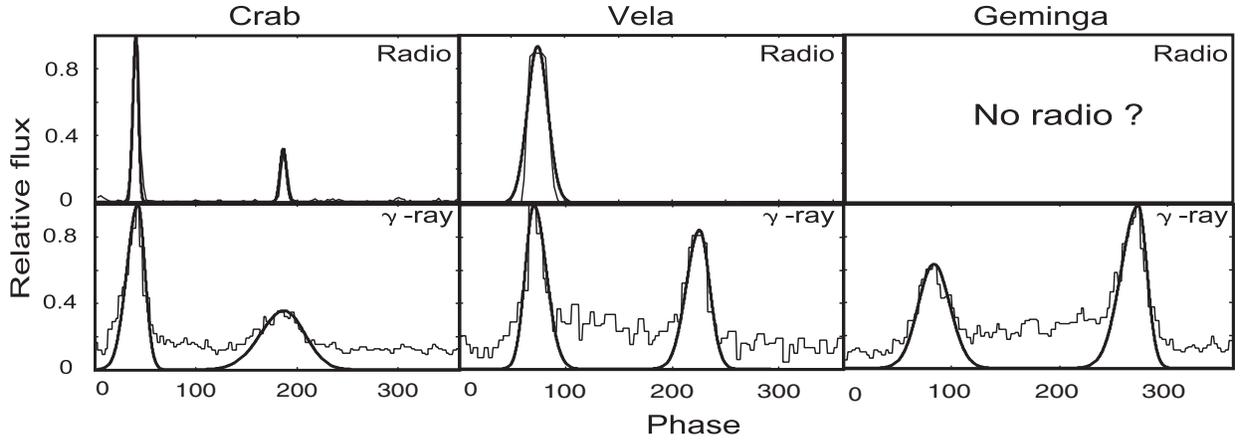} \caption{The theoretical and observed Crab-like,
Vela-like and Geminga-like light curve at both radio and
$\gamma$-ray bands. The $\gamma$-ray and radio data are derived
from Thompson 2003 and the European Pulsar Network Data Archive
respectively. The parameters listed in the figure are just for
reference. If phase separation is determined with 10\% error, we
found that $\alpha\in[36,49], \zeta\in[39,57],
\kappa\in[0.8,0.99],\lambda\in[0.65,0.85]$ for Crab and
$\alpha\in[33,44], \zeta\in[51,61],
\kappa\in[0.65,0.83],\lambda\in[0.77,0.99]$ for Vela. We do not
give parameters for Geminga light curve, the parameters can not
get only from $\gamma$-ray light curve. } \label{fit}
\end{figure}

For the last open field line, the angle between the radiation
direction and the magnetic axis, $\theta_{\rm \mu}$, is also a
function of the azimuthal angle $\phi$, which reads
\begin{equation}
\theta{\rm \mu}=\arctan\left[\frac{3 \sin(2 \theta)}{1+3\cos(2
\theta)}\right] \label{thetamu},
\end{equation}
where $\theta(\phi)=\arcsin[r(\phi)/R_{\rm 0}(\phi)]^{0.5}$. The
aberration correction is performed on the $\theta_{\rm \mu}$ by
numerical procedures. Pulse phase separations can be figured out
by direct but lengthy mathematic, which involves the impact angle
$\xi$ ($\xi=\zeta-\alpha$). The calculated pulse profiles and
parameters are given in the Fig.~\ref{fit}. The width and height
of each component in the mean pulse profile are input free
parameters. But phase separation are firmly in relation to the
parameters $\alpha$, $\xi$, $\lambda$ and $\kappa$. The physical
considerations for the parameters are as follows.

{\em Physical and geometrical limitation for $\gamma$-ray emission
regions}. The Lorentz factor of the secondary particles out of the
inner gap can reach to $10^3$ \citep{Z97}. When the secondary
moving out from the IAG, they lose their energy through various
mechanisms simultaneously, where the inverse Compton scattering
plays an important role \citep{XQWH85,DR00,ZH00}. Magnetic inverse
Compton scattering is also an important source of hard gamma-ray
photons \citep{S95}. This means that the hard $\gamma$-ray
radiation could be produced just out of the gap. However, the
position where such $\gamma$-rays can escape is limited by the
$\gamma$-ray attenuation effect in the strong magnetic fields.

A spectral cutoff of $\gamma$-ray pulsar above 10 GeV to 100 GeV
\citep{K03} presents a lower limit of the distance from the
emission location from the stellar center. Detailed calculation
shows that the 100 GeV photons can only escape from the distance
at least above 20$R$ for both the Crab (surface magnetic strength
$B_{0}$ is $3.7\times10^{12}$ Gauss) and the Vela
($B_{0}=3.3\times10^{12}$ Gauss) pulsars. We obtain $\kappa>0.5$
and $\kappa>0.2$, respectively, for the Crab and Vela pulsars in
the model.

The calculation indicates that the parameters of both $\lambda$
and $\kappa$ will affect the phase separation between two peaks of
the $\gamma$-ray light curve. For parameters of Crab and Vela (see
Fig.~\ref{fit}), we find $\kappa \in [0.8,0.99]$ and $\lambda \in
[0.65,0.85]$ for Crab, and $\kappa \in [0.65,0.83]$ and $\lambda
\in [0.77,0.99]$ for Vela, respectively, for 10\% change of phase
separation of two peaks. The values of $\kappa$ and $\lambda$
indicate that the main radiation region is confined near the NCS.

{\em Radio emission}. The ICS model \citep{Q98,X00,Q01,Q02} is
involved in our IAG model to account for the radio emission. The
secondary pairs streaming out from the polar cap cascade with
typical energy $\gamma=(1-\beta^2)^{-0.5}\sim 10^3$ will scatter
the low frequency waves produced in sparking, and the up-scattered
frequency reads $\nu \simeq 2\gamma ^2\nu_{\rm
0}(1-\beta\cos\theta _{\rm i})$ (for $B \ll B_{\rm q}=4.414\times
10^{13}$ Gauss), where $\theta_{\rm i}$ is the incident angle (the
angle between the moving direction of the particle and the
incoming photon).

The differences between the observed radio and $\gamma$-ray pulse
profiles are very significant.
Fig.~\ref{fit} shows the calculated Crab-like, Vela-like and
Geminga-like light curve for both radio and $\gamma$-ray band
under reasonable parameters. The observed pulse profiles in
gamma-ray and radio bands are reproduced for seven pulsars: the
Crab, PSR B1509-58, the Vela, PSR B1706-44, PSR B1951+32, the
Geminga (no observed radio emission) and PSR B1055-52
\citep{Q03a,Q03b}.

\section{Conclusions and discussions}

We emphasize here that the IAG and the NCS play an important role
in pulsar $\gamma$-ray radiation. At least the IAG can be a source
of high energy particles for pulsars emission. If the radiation
region locates near the NCS, many radio and $\gamma$-ray
observational facts can be easily understood. The large phase
separation of components of $\gamma$-ray light curves and the
relative compact radio phase separation reveal that the
$\gamma$-ray radiation region should locate near the NCS. The
geometry calculation partly supports this radiation location.

It is indicated from the calculation above that the condition of
$\gamma$-ray pulsars which we observed is that the view angle
$\zeta$ should be large than the inclination angle, i.e.
$\beta>0$. This is because that the part of IAG between the
magnetic axis and the equator is wider so that it gain higher
potential drop, when the inclination angle becomes larger.

The luminosity of ${\gamma}$-ray pulsars in the IAG model depends
on the maximum potential drop across the gap, $\Delta V=0.5\Omega
B r_{\rm p}^2c^{-1}$. For a given  Lorentz factor we find that the
$\gamma$-ray luminosity are proportional to ${\dot P}^{0.5}
P^{-1.5}$, which is in agreement with observations.

Our IAG scenario is intrinsically different from the free-flowing
type of polar cap model \citep{MH03}. How does the IAG interact
with the outer gap? Are pulsars bare strange stars or special kind
of neutron stars \cite{G02}? Does the radiation come from the
re-accelerated non-stationary pair flow or free flow? All these
questions need to be investigated further more.
The details of the IAG model, such as the re-acceleration and
radiation processes, $\gamma$-ray luminosity, spectral behaviors,
emission beam properties and so on, will be discussed in separated
papers.

\acknowledgements We are very grateful to Professor R. N.
Manchester and Dr. B. Zhang for their valuable suggestions.  This
work is supported by NSF of China(10373002, 10273001).

\end{document}